\documentclass[11pt]{article}

\usepackage{geometry}
\usepackage{float}
\usepackage{graphicx}
\usepackage{epstopdf}
\usepackage{natbib}
\usepackage{color}
\usepackage{bm}
\usepackage{psfrag}
\usepackage{array}
\usepackage{multirow}
\usepackage{setspace}
\usepackage[percent]{overpic}
\usepackage{amsfonts}
\usepackage{amsmath}
\usepackage{amssymb}
\usepackage{booktabs}
\usepackage{cases}
\usepackage{caption}
\usepackage{authblk}
\usepackage{verbatim}
\usepackage[small,compact]{titlesec}
\usepackage{url}
\usepackage[version=4]{mhchem}
\usepackage{textcomp}
\usepackage{flafter}

\newcommand{\ud}{\mathrm{d}}
\newcommand{\ex}{\mathrm{exp}}
\newcommand{\co}{CO$_2$ }
\newcommand{\upth}{$^{\textrm{th}}$ }

\newcommand\pfl[1]{{\color{black}#1}}
\newcommand\edit[1]{{\color{black}#1}}
\newcommand\resp[1]{{\color{black}#1}}

\begin{document}

\title{Predictive and retrospective modelling of airborne infection risk using monitored carbon dioxide}
\author{Henry C. Burridge$^{\ast,1}$, Shiwei Fan$^{2}$, Roderic L. Jones$^{2}$, Catherine J. Noakes$^{3}$, and P. F. Linden$^{4}$}
\affil{$^{1}$ Department of Civil and Environmental Engineering, Imperial College London, Skempton Building, South Kensington Campus, London SW7 2AZ, UK.\\ $^{2}$ Department of Chemistry, University of Cambridge, Lensfield Road, Cambridge CB2 1EW, UK. \\ $^{3}$ School of Civil Engineering, University of Leeds, Leeds, LS2 9JT, UK. \\ $^{4}$ Department of Applied Mathematics and Theoretical Physics, University of Cambridge, Centre for Mathematical Sciences, Wilberforce Road, Cambridge CB3 0WA, UK. \\$^{\ast}$h.burridge@imperial.ac.uk }

\maketitle

\pagestyle{plain}

\begin{abstract}

\edit{The risk of long range, herein `airborne', infection needs to be better understood and is especially urgent during the current COVID-19 pandemic}. We present a method to determine the relative risk of airborne transmission that can be readily deployed with either modelled or monitored CO$_2$ data and occupancy levels within an indoor space. For spaces regularly, or consistently, occupied by the same group of people, e.g. an open-plan office or a school classroom, we establish protocols to assess the absolute risk of airborne infection of this regular attendance at work or school. We present a methodology to easily calculate the expected number of secondary infections arising from a regular attendee becoming infectious and remaining pre/asymptomatic within these spaces. We demonstrate our model by calculating risks for both a modelled open-plan office and by using monitored data recorded within a small naturally ventilated office. \resp{In addition, by inferring ventilation rates from monitored CO$_2$ we show that estimates of airborne infection can be accurately reconstructed; thereby offering scope for more informed retrospective modelling should outbreaks occur in spaces where CO$_2$ is monitored.} Our modelling suggests that regular attendance \resp{at an office for} work is unlikely to significantly contribute to the pandemic \resp{but only} if relatively quiet desk-based work is carried out in the presence of adequate ventilation \resp{(i.e. at least 10\,l/s/p following UK guidance)}, appropriate hygiene controls, distancing measures\resp{, and that all commuting presents minimal infection risk. Crucially, modelling even moderate changes to the conditions within the office, or basing estimates for the infectivity of the SARS-CoV-2 variant B1.1.7 current data, typically results in the prediction that for a single infector within the office the airborne route alone gives rises to more than one secondary infection.}

\end{abstract}

\noindent Keywords: Infection modelling, Airborne infection risk, Monitored CO$_2$, R-number, COVID-19, Coronavirus SAR-CoV-2.
    
\noindent \textbf{Practical implications} Our readily deployable model enables the risk of infection from airborne diseases, including COVID-19, to be robustly assessed from either monitored \co or modelled data. This contribution, enables calculations of an expected number of secondary infections arising within indoor spaces regularly attended by the same/similar group of people, e.g. school classrooms and open-plan offices. Finally, we note that our model can be generically applied to all infectious airborne diseases and consideration should be given to the application of our model to indoor beyond those regularly attended spaces.

\section{Introduction} \label{sec:intro}

The novel coronavirus disease (COVID-19), which causes respiratory symptoms, was declared a pandemic by the World Health Organization (WHO) on the 11$^{\textrm{th}}$ March 2020 --- thereby marking its global impact. Transmission of such respiratory infections occurs via virus-laden particles (in this case the virus SARS-CoV-2) formed in the respiratory tract of an infected person and spread to other humans, primarily, via three routes: the droplet (or spray) route, the contact (or touch) route and the airborne (or aerosol) route \citep[e.g. see][]{Mittal20,Milton20,Li2021}. According to the WHO, ``Airborne transmission is defined as the spread of an infectious agent caused by the dissemination of droplet nuclei (aerosols) that remain infectious when suspended in air over long distances and time'' \citep{WHO14}. After some initial resistance, and significant pressure from the scientific community \citep[e.g.][]{MORAWSKA2020105832,Morawska20time}, the WHO finally acknowledged the possibility of airborne infection for COVID-19 on the 8$^{\textrm{th}}$ July 2020 \citep{Indep}. \resp{In the latter part of 2020 multiple mutations to the SAR-CoV-2 virus conspired to give rise to a new variant, named B1.1.7 \citep[see, for example,][for a more detailed discussion]{Kupferschmidt21}. This variant is currently thought to be significantly more infectious than pre-existing strains \citep{SAGE20} and is increasingly prevalent within the UK and other parts of Europe. At the start of 2021, with new more successful variants arising, COVID-19 infection levels remain worryingly high around much of the world.} In our present article, we focus on assessing the risk of infection of respiratory diseases via the airborne route, taking COVID-19 as an example; ultimately, deriving a methodology for calculating the expected number of secondary infections that might arise within any indoor space that is regularly attended by the same group of people, applicable to any airborne disease \resp{(with estimates for the duration over which infectors remain pre/asymptomatic.)} \resp{We comment on the airborne infection risk for COVID-19 within open-plans offices under a variety of environmental conditions and include consideration of the variant B1.1.7.}

The pioneering work of \citet{Wells55} and that which followed by \citet{Riley78} established methods, commonly referred to as the Wells-Riley model, for quantifying the risk of airborne infection of respiratory diseases. \edit{Unlike dose-response models, which assess the likely infection response to some (frequently cumulative) dose, Wells-Riley models typically report the complementary probability that no-one becomes infected. As such, these models do not rely on assessing the cumulative exposure which could prove problematic when assessing the infection risk over durations of varied occupancy. Early formulations \citep[e.g.][]{Riley78} were restricted to indoor spaces which were in a steady-state with a known constant rate of ventilation of outdoor air. The requirement of steady-state is avoided by the formulation of the model presented by \citet{Gammaitoni97}.} \citet{Rudnick03} further extended the practical application of the Wells-Riley model by negating the need to assess nor assume the rate of ventilation of outdoor air -- a notoriously difficult quantity to measure directly (see {Appendix} \ref{app:Q} for a detailed discussion). \citet{Rudnick03} achieved this via the realisation that the risk of airborne infection could be directly inferred via measurements of \co ``if the airspace is well mixed''. We generalise the model of \citet{Rudnick03} relaxing the assumption of a well-mixed space and to further account for occupation profiles and activity levels which vary in time. 

For many airborne infections the likelihood of spread within the vast majority of indoor spaces, even over periods of a few hours, is reasonably low (as we show for the coronavirus SARS-CoV-2 and the resulting disease COVID-19). However, we suggest that there exists a significant proportion of indoor spaces which are, for the majority of each working day, attended by the same/similar group of people (e.g. open-plan offices and school classrooms), herein `regularly attend spaces'. Our model enables the likelihood of the spread of infection via the airborne route to be calculated (from either easily obtainable monitored data or modelled data) over multiple day-long durations. Hence, in the case of COVID-19 (for which infectors are estimated to remain pre-asymptomatic for 5-7 days) our model calculates the likely number of people that become infected during a period in which a pre/asymptomatic infector regularly attends the space.

We derive an extended airborne risk model in \S\ref{sec:model}, assess the risk of infection in a modelled open-plan office in {\S\ref{sec:quanta}}, and use monitored data from naturally ventilated office {in} \S\ref{sec:res} to estimate the infection risk. \resp{We describe retrospective modelling of an office in \S\ref{sec:resRet}, and we} draw conclusions {in} \S\ref{sec:conc}. 

\subsection{The Wells-Riley approach to airborne infection risk}

The pioneering work of \citet{Riley78} defined the infectivity rate as
\begin{equation}
    \lambda = \frac{I \, p \, q}{Q} \, ,
\end{equation}
where $I$ is the number of infected people, $p$ is the breathing (pulmonary ventilation) rate, $Q$ is the ventilation (outdoor air supply) rate, and $q$ is the unit of infection, quantum \citep[see][for discussion]{Riley78}, which varies significantly between disease, with activity level, and (as with all biologically derived parameters) with individual human beings. For {many} diseases and relevant activity levels, appropriate values of $q$ have {been} determined and {are} reported {in the} literature \citep[e.g.][]{Noakes12,Bueno20Q,Miller20} -- however, significant uncertainties are associated with these values. Moreover, the variability due to individuality is challenging to reflect, see {\S}\ref{sec:quanta} for more discussion. In particular, high values of risk are obtained {from }quanta generation rates derived from so called `superspreading events' \citep[e.g. see][]{Miller20} --- we choose not to focus on such cases but note that should we have done so then the risks reported herein would be dramatically increased (see, for example, table \ref{tab:R}). For a given demographic and activity level within the space the breathing rate $p$ can be taken as constant and values are widely reported in the literature, the number of infected people $I$ is an input to the model usually taken to be constant.

\citet{Riley78} were no doubt aware of the significant challenges in measuring, or even inferring, the outdoor air supply rate to a given indoor space (see {Appendix} \ref{app:Q}). 
Instead, it was chosen to report the model in a form that can only be applied to indoor spaces for which the air is relatively well-mixed and the flows are in steady-state. Under these restrictive assumptions the classical Wells-Riley equation is recovered, namely that the likelihood, $P$, that infection spreads within a given indoor space during a time interval $T$ is \begin{equation}
    P = 1 - \exp{\left( - \frac{I \, p \, q}{Q} T \right) } \, . \label{eq:P1}
\end{equation}

\section{A model for airborne infection risk in transient spaces with variable occupancy and activity levels} \label{sec:model}

\edit{We present a simple model to estimate airborne infection risk capable of both exploiting data of the environmental conditions concerning the ventilation (i.e. \co measurements), and accounting for occupancy levels that vary in time. As t}he insightful work of \citet{Rudnick03} highlighted, airborne infection can only occur through the breathing of rebreathed air that is infected. 
\edit{It is important to note that respiratory activity, e.g. breathing, results in a complex multi-phase flow being exhaled. Relative to inhaled air, exhaled air is typically warmer, of higher moisture content (both in the form of vapours and droplets), richer in \co\!, and contains more bioaerosols of which some may be viral particles. The fate of viral particles is of particular relevance to estimating infection risk, and determines via which of the three routes (droplet, contact or airborne) infection might occur, see \citet{Milton20,Li2021}. Those virus particles held in larger droplets may either be directly sprayed onto another individual (risking transmission via the droplet route) or fall to surfaces (potentially giving rise to transmission via the contact route). Following the \citet{WHO14} definition, virus particles that might give rise to transmission via the airborne route must remain ``suspended in air over long distances and time''. Therefore, the viral aerosols that can give rise to airborne infection (transmission via the airborne route) will be largely carried with the gaseous emissions exhaled. Directly detecting the presence of viral aerosols within air is challenging, costly and impractical to implement at scale. However, gaseous emissions exhaled \pfl{by persons} are relatively rich in \co\!, and  \co sensors of suitable accuracy (say $\pm 50$\,ppm) are easily obtained for a moderate cost. Hence monitoring \co as a proxy for air that has the potential to be carrying viral aerosols, whilst being far from a perfect tracer, has not only legitimate scientific grounds but is also practical to implement at scale.}

Within most indoor spaces human breathing is the dominant source \co and so the fraction {$f$} of rebreathed air can be inferred from the ratio of the \co concentration within the space (above outdoor levels, $C_0$) to the concentration of \co added to exhaled breath during breathing, $C_a$, giving
\begin{equation}
    f = \frac{C - C_0}{C_a} \, ,
\end{equation}
where $C$ in the measured \co within the space. Denoting the number of people {$n$} within the space \resp{and taking the occupancy to be constant} gives the fraction of rebreathed air that is infected as $f \, I/n$. \edit{Note that estimating the likelihood of airborne infection via monitored CO$_2$, from which the fraction of infected rebreathed air is estimated \citep[e.g.][]{Rudnick03}, makes no stronger assumptions than are already implicit within the classical formulation of the Wells-Riley equation (\ref{eq:P1}). Within the Wells-Riley equation the ratio, $(I\,p)/Q$, is the fraction of infected air estimated to be within the space while the formulation of \citet{Rudnick03} expresses it as $(f \, I)/n$. In both formulations the estimate of the fraction of infected air is translated into a likelihood of infection rate via the quanta generation rate. \resp{The quanta generation rates utilised i}n most studies (including the present study) are deduced from data concerning actual far-field infection events. \resp{Obviously, these} infection events occur as a result of the full physics governing the complex transport of viral particles. As such, \resp{the empirical data (underlying the estimates of quanta generation rates)} implicitly accounts for some of the differing physics \resp{expected to} arise between the transport of hypothesised gaseous infectious air (required by Wells-Riley based models) and the actual transport of infectious particles \resp{--- which for airborne infection to occur must (by the definition of the transmission route)} be able to be `suspended in air over long distances and time'.}

\citet{Rudnick03} chose to express their result as
\begin{equation}
    P = 1 - \exp{\left( - \frac{I}{n} q \int_0^T f \, \ud t \right) } = 1 - \exp{\left( - \frac{I}{n} q \, \overline{f} \, T \right) } \, . \label{eq:PRM}
\end{equation}
As they point out, this result ``has very general applicability; it is valid for both steady-state and non-steady-state conditions and when the outdoor air supply rate varies with time''. \edit{Furthermore, we highlight that their assumption of a well-mixed space is unnecessary. The fraction of rebreathed air, $f$, is based on a point measurement of \co which, assuming human respiration is the dominant source of \co (entirely reasonable in the absence of other sources, e.g. unvented combustion), provides, at any instant, a good estimate of the fraction of air at that point within the space that has already been breathed by another individual. This point measurement can be integrated according to (\ref{eq:PRM}) to give the likelihood that a person (at the same location as the \co sensor within the space) becomes infected assuming only that the infected air is relatively well-mixed within the uninfected rebreathed air, i.e. it does not require that all the air within the space is well-mixed. This implies that, where multiple \co sensors within a single indoor space (inevitably) give different readings, airborne infection risk can be assessed without violation of the modelling assumption being implied; in fact, the different readings within the space could be exploited to obtain estimates of the spatial variation in risk. }

\edit{We wish to extend the generality of (\ref{eq:PRM}) with greater application in mind, in particular to account for occupancy levels that vary in time (see \S\ref{meth:R}). The likelihood \resp{that airborne infection occurs} within a given space can be determined from
\begin{equation}
    P = 1 - \exp{\left( - \int_0^T \lambda \, \ud t \right) } = 1 - \exp{\left( - \int_0^T \sigma(n) \; f_i  \, q \; \ud t \right) } \, , \label{eq:Pgen}
\end{equation}
with, in the general case, the fraction of infected air $f_i$ within the space being determined by solution of
\begin{equation}
    \frac{\ud f_i}{\ud t} = \frac{I \, p}{V} - \frac{f_i \, Q}{V} \, , \label{eq:figen}
\end{equation}
where $V$ is the volume of the indoor space which is typically easily estimated, and \resp{$\sigma(n)$ is determined based on whether the space is occupied or unoccupied: $\sigma(n)=1$ for $n>0$ and $\sigma(0)=0$}. For a derivation of (\ref{eq:Pgen}) from first principles see Appendix \ref{app:first}. With suitable selection of the time at which to initiate investigation, the initial condition $f_i(0) = 0$ will frequently be suitable; typically values for the breathing (pulmonary ventilation) rate $p$ can be sourced from the literature, and from monitored \co and occupancy levels the ventilation rate $Q$ can be estimated from (\ref{eq:Q}), although as discussed in Appendix \ref{app:Q} such estimates are subject to considerable noise. In spite of this noise, when \co is monitored and occupancy levels are available then solution of (\ref{eq:Pgen}) is only lacking knowledge of the time series of the number of infectors $I$ within the space. Note that solution of the full system of equations, e.g. (\ref{eq:Q}), (\ref{eq:Pgen}) and(\ref{eq:figen}), does require the assumption that the air within the space be approximated as well-mixed because one is required to calculate the ventilation rates explicitly; unlike simpler models e.g. (\ref{eq:PRM}) and (\ref{eq:Pus}) which do not. Where retrospective modelling is being undertaken to assess a particular outbreak, or spreading event, estimates of these data may be available and attempts to apply the above might prove useful and we turn our attention to informing these cases in \S\ref{sec:resRet}. However, with a focus on predictive modelling which, by definition, requires some assumption regarding the presence of infectors $I(t)$, we now consider some appropriate assumptions.

We will either assume that occupants arrive and leave over realistic periods of time (i.e. we model them to not all arrive and leave at once), or our monitored data show this to be so, and thus there exists at least two reasonable principles by which to establish the presence of infectors, $I(t)$. \pfl{At one extreme,} assume that the infector is always the first to arrive and the last to leave. Alternatively, one could assume that there is always a constant proportion of the (current) occupants infected such that when the space is occupied to design capacity there is a single infector (this results in the number of infectors, $I$, taking non-integer values outside full design occupancy which is inconsequential). Should one choose to assume the former, there is potential that risks are over estimated reported for scenarios in which occupancy is decreased, or equivalently by allowing more occupants one could under report the risk of the space since a lesser proportion of the occupants are infected --- in the absence of knowledge as to who is infected, this cannot be reasonable for comparison of risk with different occupancy levels. As such, we choose to make the latter assumption, i.e. there is always a constant proportion, $\alpha$, of the (current) occupants infected, i.e. $I(t) = \alpha \, n(t)$. Doing so renders $I(t)/n(t) = \alpha$ as constant or, alternatively, that the proportion \pfl{$\alpha$} of rebreathed air that is infected remains constant, and as such we can write the likelihood of airborne infection in a far simpler form that still allows for variable occupancy, namely
\begin{equation}
    P = 1 - \exp{\left( - \int_0^T \sigma(n) \; f  \frac{I}{n} q \; \ud t \right) }  \, , \label{eq:Pus}
\end{equation}
where we choose to include presentation with the fraction $I/n$ within the integral to emphasise that both the numerator and the denominator vary in time. Making the assumptions that result in (\ref{eq:Pus}) being valid not only seems reasonable for predictive modelling but it enables the work of \citet{Rudnick03} to be extended to predict the airborne infection risk, based on either modelled or monitored \co data, within indoor spaces that also have variable occupancy levels, crucially, without need to solve the more general equation (\ref{eq:Pgen}). To generally solve (\ref{eq:Pgen}), on is required to solve (\ref{eq:Q}) to estimate the volume flux as an input to then enable solution of (\ref{eq:figen}); doing so requires monitored occupancy levels, \co levels, and relies on the \resp{(temporal)} gradients in \resp{monitored} \co levels. Presented in the form (\ref{eq:Pus}) and noting $I(t)/n(t) = \alpha$, highlights that precise occupancy levels are not required for the assessment of infection risk, only knowledge of the occupied periods is required. 

Throughout this study we choose to set $\alpha$ such that there is a single infector present, i.e. $I(t)=1$, when the space is occupied to design capacity, $N_d$, which gives $\alpha = I(t)/n(t) = 1/N_d$. Doing so is as reasonable choice as any other but we note that should comparison of airborne infection risk between separate indoor spaces of differing design capacity be desired then alternate choices should be made. The applicability of the choices to enable predictive modelling via (\ref{eq:Pus}) will be highlighted throughout \S \ref{sec:res}, and we return to solving the more general equations in \S\ref{sec:resRet}.}


\subsection{Quantifying the relative risk for changes in environmental management}

To examine the effects of a particular change in conditions
within a given indoor space, e.g. change in ventilation rate, occupancy level/behaviour, etc., it is informative to define a `base case' scenario for which the likelihood of infection during a time interval $T$ is $P_0$ and quantify the airborne infection risk of chosen scenarios relative to the base case. \edit{We can then define the risk of some test scenario relative to the base case as $RR = P/P_0$, denoting this relative risk as $RR_A$ when a pre/asymptomatic is investigated, i.e. $T=T_A$. It is worth noting that the relative risk can be written as a ratio of Taylor series expansions of the exponential terms. Doing so can aid approximation since, when the integral $\int_0^{T_s} \sigma f  \, \frac{I}{n} \, q \; \ud t$, is small, the leading order terms in the expansion dominate and the relative risk ceases to be dependent on the quanta generation rate --- a notoriously difficult quantity to parameterise which also varies widely between diseases. Hence for airborne infection risk assessment the leading order expansion for the relative risk can be reported as valid for all diseases (note that the duration $T$ for which the approximation remains valid does change with disease). Moreover, for any given disease results for the relative risk can be reported with a greater degree of certainty, irrespective of the duration.} 

\subsection{Defining absolute risk and the expected number of secondary infections for a given indoor space} \label{meth:R}

An indoor space can be considered as contributing to the spread of a disease if an infected person attends the space for a duration over which it is more likely than not that they infect others. In the case that someone is showing symptoms of the disease it is reasonable to assume that they
cease attending the space or that they be required to do so. Individuals can remain infectious and asymptomatic/presymptomatic for time periods of multiple days (which we denote as $T_A$) and this renders (\ref{eq:PRM}) unsuitable for quantifying this likelihood for most indoor spaces. However, for  regularly attended {spaces} 
e.g. open plan offices and school classrooms, the probability $P_A$ that someone becomes infected via the airborne transmission route (assuming an infected person attends the space) can be robustly determined via our formulation (\ref{eq:Pus}). To do so, time series data for the rebreathed air fraction (monitored or modelled), the occupancy level and quanta generation rate are required over the duration $T_A$. For a given disease, assuming the activity levels (per capita) remain broadly {the same} within the space, the quanta generation rate can be assumed constant. For real-world assessment, $f$ and $n$ can be obtained from monitored \co and occupancy data, respectively. Moreover, for model cases this can easily be calculated. We demonstrate examples of this for model building spaces (\S\ref{sec:resMod}), and using monitored data an existing open plan office (\S\ref{sec:resOff}) taking COVID-19 as a case study. 

As elegantly pointed out by \citet{Rudnick03} their formulation (\ref{eq:PRM}) can be used to determine \edit{what they term a `basic reproductive number'} for an airborne infectious disease within an indoor space. \edit{Herein, we describe this as the expected number of secondary infections via the airborne route,  $S_I$, that arise within an indoor space} when an infectious individual is attending the space and everyone else is susceptible. \edit{For regularly attended spaces, this is simply calculated from the probability of someone becoming infected over the pre/asymptomatic period multiplied by the number of susceptible people, giving}  
\begin{equation}
    S_I = (N_a-1) \left[ 1 - \exp{\left( - \int_0^{T_A} \sigma(n) f  \frac{I}{n} q \; \ud t \right) } \right] \, , \label{eq:R}
\end{equation}
where $N_a$ is the total number of people that regularly attend the space. We earlier pointed out that calculations of the likelihood of infection via either (\ref{eq:PRM}) or via (\ref{eq:Pus}) do not require the assumption that all the air within the indoor space is well-mixed. However, for the expected number of secondary infections to be a meaningful estimate then (\ref{eq:R}) requires that the likelihood of infection to be representative of the risk through the occupied indoor space. If all the air within the space is well-mixed then this is simply satisfied; otherwise, multiple \co measurements should be taken, within the breathing zone, to assess the degree of variation.

\vspace{\baselineskip}
To summarise our modelling, we have developed practical statistics to assess airborne infection via relative risk based scenario testing ($RR$), the absolute probability of infection ($P_A$), and the expected number of secondary infections for an indoor space ($S_I$). All of these can be calculated by obtaining/modelling representative \co data. Moreover, for measured/modelled \co distributions within the space, on assuming the infected \& uninfected rebreathed air are mixed, these statistics can be calculated and their variation within the space investigated.

\section{Determining appropriate quanta generation rates} \label{sec:quanta}

\edit{As with all Wells-Riley based infection modelling an input parameter for which great uncertainty abounds is the quanta generation rate, $q$ --- with the novelty of COVID-19 this uncertainty in compounded. Given the uncertainty, we include results of scenario tests at various feasible levels of $q$, which span nearly four orders of magnitude based \pfl{on} the data of \citet{Buonanno20}}. As a base case, which we deem appropriate for the regularly attended spaces on which we focus (namely, open-plan offices and class rooms) we take a value of $q=1$\,quanta/hr --- this is obtained by taking ${c_x = c_i \, c_v \approx 7 \times 10^{6}}$\,RNA/ml, where $c_i = \{0.1, 0.01\}$ is the ratio between infectious quantum and the infectious dose expressed in viral RNA copies, and $c_v = \{7 \times 10^7, 7 \times 10^8\}$\,RNA/ml is the viral load measured in sputum. These values obtained by consideration that for most of the time, in most open-plan offices and classrooms, most of the occupants are sitting breathing with perhaps a small number vocalising --- the data for whispered counting falls between these two activities and is rather more close to breathing --- as such, for our base case, we take data for whispered counting from \citet{Buonanno20} and use their results to map our selected values of $c_x$ to values of quanta generation rates $q$. Moreover, we consider a scenario in which the occupants within the open-plan office or classroom are (on average) all vocalising/talking (e.g. a call-centre or noisy classroom), taking again $c_x \approx 7 \times 10^{6}$\,RNA/ml gives $q \approx 5$\,quanta/hr. In addition, we consider a scenario in which the viral load in sputum is somewhat reduced, i.e. $c_v \approx \{2 \times 10^7, 2 \times 10^8\}$\,RNA/ml, giving $q \approx 0.3$\,quanta/hr.

 \resp{Mutations of the SARS-CoV-2 virus have given rise to a new variant, named B.1.1.7, which has become widely detected in certain geographical regions (within the UK in particular) in the latter part of 2020 \citep[see][for details]{Kupferschmidt21}. This variant, which has already spread across international borders, is believed to be potentially around 70\% more transmissible than the, herein, `pre-existing' strains of the virus \citep{SAGE20,Tang20}. It is, as yet, unclear by which mechanisms the transmission of the new variant is increased; however, it is an important development which demands analysis. We therefore include estimates for the airborne infection of variant B1.1.7 within our results. To do so, we assume the increase in transmission of B1.1.7 via the airborne route might be proportional to the total increase and, for the various scenarios considered, take quanta generation rates for the variant, $q_v$, to be 70\% higher than those corresponding to pre-existing strains of the virus, i.e. $q_v = 1.7q$.}

\section{Predictive modelling \edit{of airborne infection} using COVID-19 as an example} \label{sec:res}

\subsection{Application to a model open-plan office} \label{sec:resMod}

By way of example, we first consider a moderately sized open-plan office, of floor area 400\,m$^2$ and \pfl{(a generous)} floor-to-ceiling height 3.5\,m, which is designed to be occupied by 40 people \citep[][]{CIBSEA}. We assume that occupants arrive steadily between 08:00 and 09:00 each morning, each take a one hour lunch break during which they leave the office, and leave steadily between 17:00 and 18:00 each day. While within the office we assume that (on average) each occupant breathes at a rate of approximately $p=8$\,l/min with a \co production rate of 0.3\,l/min, giving $C_a = 0.038$ and we take the outdoor \co level to be 400\,ppm \citep{Rudnick03}. \edit{As a base case we assume ventilation provision inline with UK guidance for office spaces, i.e. $Q_{pp} = 10$\,l/s/p \citep{CIBSEA}, or a total ventilation rate of $Q = 400$\,l/s.}

Our model run for this open-plan office gives, for the base case, the absolute risk of infection during a period of pre/asymptomatic COVID-19 infection as $P_A = 1.1\%$. If one had have taken the classical Wells-Riley model (\ref{eq:P1}), taking there to always be a single infector present and $T$ to be the simple sum of occupied hours (i.e. $T=40$\,hrs), the level of risk reported would have been $P = 1.3\%$, around 20\% higher. We note the key benefit of our model is the ability to use monitored \co and occupancy data as we show in \S\ref{sec:resOff}.

\subsubsection{The impact of varied quanta generation rates} \label{sec:resq}

\begin{figure}
             \begin{center}
             \includegraphics[width=0.95\textwidth]{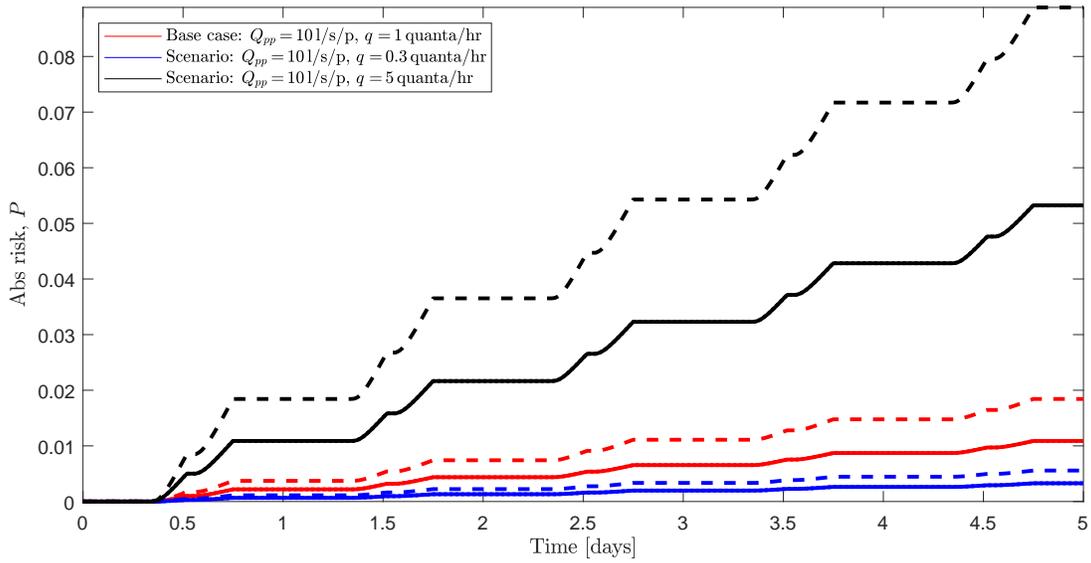}
             \end{center}
             \caption{The variation in the likelihood of infection with time over the five day pre/asymptomatic period \resp{with ventilation inline with UK guidance, i.e. $Q_{pp}=10$\,l/s/p. Solid curves mark the risk with differing quanta generation rates assumed for the pre-existing SARS-CoV-2 strains: blue denotes $q=0.3$\,quanta/hr, red denotes $q=1$\,quanta/hr, and black denotes $q=5$\,quanta/hr. The correspondingly coloured dashed curves mark estimates for the variant B1.1.7 for which we take the quanta generation rates to be $q_v=1.7q$.}}\label{fig:Mod_Pabs_q}
\end{figure}

We first examine the impact of varied quanta generation levels; namely, $q=\{1, 0.3, 5\}$ (see \S\ref{sec:quanta} for a fuller discussion) \edit{on the likelihood of airborne infection}. Figure \ref{fig:Mod_Pabs_q} plots the absolute likelihood that someone becomes infected within the office \edit{via the airborne route} over the period during which an infector is expected to remain pre/asymptomatic, i.e. 5 working days (since the period of pre/asymptomaitc infectivity for COVID-19 is estimated as 5--7 days). The plot shows that in the base case the absolute risk, $P_A$, of \edit{airborne} infection within this open-plan office is just around 1.1\% (or assuming a lower viral load is appropriate the risk drops to around $P_A =0.3\%$). However, if the open-plan were a call-centre then this risk that someone becomes infection through attending work increases to above 5\%. \resp{These results correspond to estimates for the quanta generation rates based on the work of \citet{Buonanno20} on data for `pre-existing' strains of the SARS-CoV-2 virus. In the latter part of the year 2020 mutations have gave rise to a new variant of the virus B1.1.7 which is already prevalent in some geographical regions. Within figure \ref{fig:Mod_Pabs_q} dashed lines show estimates for the likelihood of airborne infection for the B1.1.7 variant based on the current data which suggest this variant may be 70\% more effectively spread. Since for the scenarios being tested the likelihoods remain relatively linear in response to changes in infectivity the risks are increased by a factor of around 1.7 in all three scenarios, i.e. the for the base case scenario with variant B1.1.7 the airborne infection is predicted to be nearly 2\%.} 

One can, of course, examine the relative risk of airborne infection; as expected from \edit{from consideration of Taylor series expansions of the exponential terms,} the results are broadly constant in time, with the relative risk taking an initial value of $q/q_0$ \resp{(or $q_v/q_0$)}, and then remaining dominated by \edit{the ratio of the quanta generation rate between the scenarios \resp{and/or variants}}. For example, in this office at the end of a pre/asymptomatic period, examining the scenario that the open-plan \edit{office changes to become} equivalent to a call-center gives the relative risk as $RR_A = 4.9$ \edit{(for $q/q_0=5$\resp{, and taking $q_v=1.7q$ gives $RR_A = 8.2$})}, and imagining that the appropriate viral load for the disease\edit{, for some reason, becomes} lower gives $RR_A = 0.3$ \edit{(for $q/q_0=0.3$)}. \edit{One can see that for these cases the relative risk is well predicted by a linear approximation (taking only the first order terms in the Taylor series expansion of the probabilities), i.e. the relative risk is approximately equal to the ratio of quanta generation rates between the scenarios.} 

\subsubsection{The importance of ventilation/outdoor air supply rates} \label{sec:resQ}

\begin{figure}
             \begin{center}
             \includegraphics[width=0.95\textwidth]{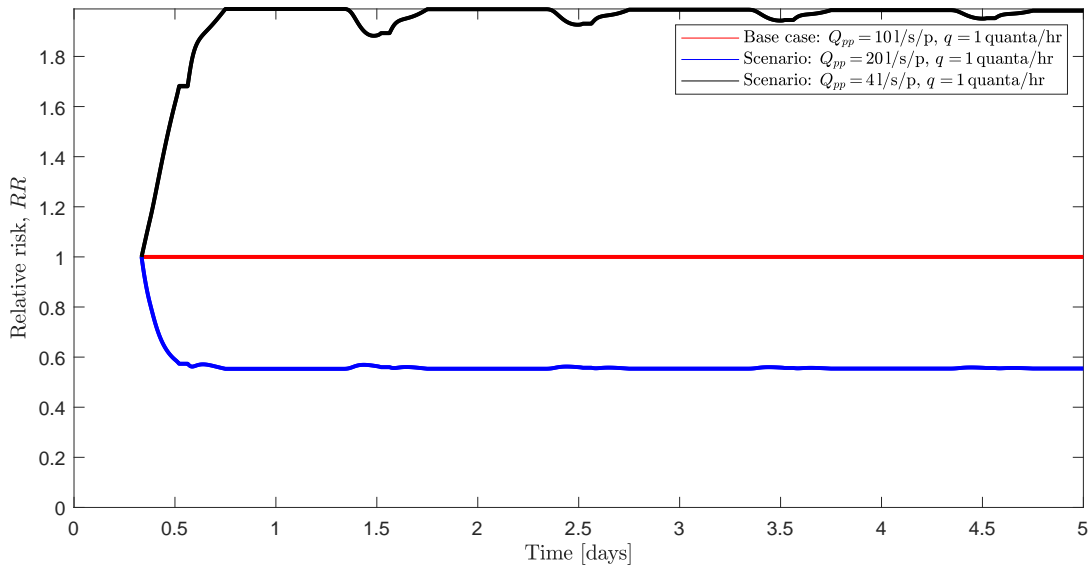}
             \end{center}
             \caption{The variation in the relative risk, $RR$, of infection with time over the five day pre/asymptomatic period. The different curves highlight different scenarios, namely: the base case, $Q_{pp}=10$\,l/s/p (red), increased vent, $Q_{pp}=20$\,l/s/p (blue), and decreased vent, $Q_{pp}=4$\,l/s/p (black). \resp{The periodic deviations from the steady-state values of $RR$ arise due to transient effects as the office is reoccupied each day.}
             }\label{fig:Mod_RR_Q}
\end{figure}

The qualitative increase in the \edit{airborne infection} risk within the office during a period of pre/asymptomatic infection for varied outdoor air supply rate per person, $Q_{pp}$, is broadly similar to that shown in figure \ref{fig:Mod_Pabs_q}. \edit{Over the full pre/asymptomatic period t}he base case (of course) again gives $P_A = 1.1\%$ \resp{for pre-existing strains of the SARS-CoV-2 virus}. Doubling the ventilation rate per person to $Q_{pp}=20$\,l/s/p \edit{decreases the likelihood to} $P_A = 0.6\%$, and decreasing the outdoor air supply rate per person to $Q_{pp}=4$\,l/s/p results in $P_A = 2.2\%$ {(see table~\ref{tab:R})}. 

Examining the relative risk $RR$, for scenarios of changing ventilation rates, then $RR$ takes an initial value of unity and only over time does the ventilation alter the accumulation of infected re-breathed air within the space. In this open-plan office\edit{, in the case that the ventilation is doubled from one scenario to the next,} the relative risk reaches an approximately steady value of $RR \approx 0.55$ around 6 hours \edit{after being occupied (see figure \ref{fig:Mod_RR_Q})}, and in the case of decreased ventilation ($Q_{pp}/Q_0 = 0.4$) then $RR \approx 2.0$ is reached after approximately 10 hours. \resp{At the start of} each \resp{working day, as the office is reoccupied, the relative risk can be seen to deviate from, then return back to, its steady-state value (figure \ref{fig:Mod_RR_Q}), this is as a consequence of transient effects --- as} the fraction rebreathed air increases at different rates in \resp{each} scenario \resp{(due to the differing ventilation rates). Transient effects that occur at the end of the day do not affect the infection risk since the office is unoccupied. These deviations lessen each day as the transients have a smaller impact on the integral risk; the deviations are also more pronounced in the case that the ventilation is reduced, \edit{i.e. the} black curve in figure \ref{fig:Mod_RR_Q}.}

\subsubsection{The expected number of secondary infections for an open-plan office} \label{sec:Rnumb}

\begin{table}[]
    \centering
    \begin{tabular}{l|c |c |c}
         Secondary infections, $S_I$ & \; $Q_{pp}=4$\,l/s/p \; & \; \boldmath$Q_{pp}=10$\,\textbf{l/s/p} \; & \; $Q_{pp}=20$\,l/s/p \;   \\
         \hline
         $q=0.3$\,quanta/hr & 0.25 & 0.13 & 0.07 \\
         \resp{Variant B1.1.7} & [0.43] & [0.22] & [0.12] \\
         \hline
         \boldmath$q=1.0$\,\textbf{quanta/hr} & 0.84 & \textbf{0.42} & 0.24  \\
         \resp{Variant B1.1.7} & [1.4] & [0.72] & [0.40] \\
         \hline
         $q=5.0$\,quanta/hr & 4.0 & 2.1 & 1.2  \\
         \resp{Variant B1.1.7} & [6.6] & [3.5] & [2.0] \\
         \hline
         \\
         $q=20$\,quanta/hr & 14 & 7.6 & 4.4  \\
         \resp{Variant B1.1.7} & [20] & [12] & [7.3] \\
         \hline
         $q=100$\,quanta/hr & 35 & 26 & 18  \\
         \resp{Variant B1.1.7} & [38] & [33] & [25] \\
    \end{tabular}
    \caption{The expected number of secondary \edit{airborne} infections, $S_I$, for COVID-19 arising within an open-plan office (floor plan of 400\,m$^2$ and floor-to-ceiling height of 3.5\,m) occupied by 40 people for 8\,hrs each day over the period that a pre/asymptomatic person remains attending work. \resp{Bold text highlights the scenario (based on quiet desk-based work) taken herein as the base case; scenarios of $q=5$\,quanta/hr are intended to be representative of of more vocal office environments, and higher quanta generation rate are intended to be indicative of superspreading scenarios. Values within square brackets provide estimates for the SARS-CoV-2 variant B1.1.7 at quanta generation rates $q_v$ relative to the pre-existing strains, i.e. $q_v=1.7q$.}}
    \label{tab:R}
\end{table}

We run our model for the expected number of secondary \edit{airborne} infections (\ref{eq:R}) for a period of pre/asymptomatic infectivity (5--7 days, i.e. spanning 5 working days) varying both the quanta generation rate (\resp{of `pre-existing' virus strains} $q=\{0.3, 1.0, 5.0\}$\,quanta/hr) and the outdoor air supply rate air supply rate ($Q_{pp}=\{4, 10, 20$\}\,l/s/p). The results are presented in table \ref{tab:R} and \edit{suggest} that, for \edit{these environmental conditions with quiet desk-based work being conducted,} it is unlikely that an employee's \resp{time within the office} will significantly contribute to the spread of COVID-19 \edit{via the airborne route}. However, if the RNA copies/viral load are as expected by \citet{Buonanno20}, i.e. $q \approx 1$ for quiet desk-based work, but the office is poorly ventilated then the expected number of secondary infections arising within this office may hover dangerously close to unity \resp{for pre-existing virus strains; should the B1.1.7 variant become prevalent then a single infection within the office could be expected to give rise to around 1.4 new infections just via the airborne route}. \edit{Alternatively}, if the office is used for particularly vocal activities, e.g. a call-center or sales office \resp{(i.e. making $q \approx 5$ more appropriate for pre-existing strains), then just through their occupation} these spaces may significantly contribute to the spread of COVID-19 \resp{via the airborne route even when ventilated inline with current UK guidance --- with the situation only worsened in the presence of the B1.1.7 variant}.

\edit{To end this section, we note that \citet{Buonanno20} report far higher quanta generation rates for `superspreaders'. The definition of a superspreader is unclear and far from unanimous. We note that the term may refer to specific combinations of the particular activity being undertaken, the environmental quality, and the biological response of individuals. 
Within table \ref{tab:R}, we include two \resp{sets of scenarios (based on $q=20$\,quanta/hr and $q=100$\,quanta/hr, respectively) which report the expected number of secondary infections that might arise within our office should some superspreading event occur within. The results are worrisome with the majority of employees becoming infected from the presence of a single infector in many of the scenarios examined. We hope that the conditions that might be required to give rise to superspreader events are unlikely to occur over durations comparable to a full pre/asymptomatic periods; if so, these scenarios may prove to be overly pessimistic. }}

\edit{
\subsubsection{The benefits of reduced occupancy} \label{sec:resOcc}

The above results suggest, under certain conditions, occupation of an open-\pfl{plan} office may contribute to the spread of COVID-19 just by the airborne route. This is troubling as the airborne route is perhaps the most difficult transmission route to mitigate against with appropriate ventilation being the primary mitigation strategy. Employers should help to mitigate the airborne spread of COVID-19 by ensuring ventilation systems are sufficient to comply with guidance and that they are properly maintained. However, large scale changes to the ventilation provision, for example to double the supply of outdoor air, are costly and will take time to implement appropriately (e.g. ensuring that the heating provision, and other factors, are also adequately adjusted or upgraded). One course of action that may be more immediately appealing is to consider keeping the occupancy reduced. For example, introducing week-in week-out working would result in the occupancy being halved. However, the ventilation system can be set to keep running at the full design capacity (which in our model office was $Q=400$\,l/s in the base case scenario). Doing so results in the expected number of secondary infections that might arise via the airborne route within our office being reduced by a factor of about four, because, all else being equal, in this example the probability of an infector being present is roughly halved compounded by the fact that there are half the number of people to infect. This fourfold reduction in secondary airborne infections is significant while the strategy provides opportunities for employees to attend the office in a manner which might be of practical benefit to themselves and their employer alike. Reducing the occupancy by a factor of $r$ results in the expected number of secondary infections that might arise via the airborne route being reduced by a factor $r^2$ for all the scenarios considered herein.
}

\subsection{Airborne infection risk from monitored data in open plan offices} \label{sec:resOff}

To demonstrate the application of our model to indoor spaces with monitored \co and occupancy data we were provided access to data recorded by the `Managing Air for Green Inner Cities (MAGIC)' project (\url{http://www.magic-air.uk}). The data were recorded in a small office which had a design capacity of eight people, although during the times for which we were provided data never more than six people attended the office. The office is naturally ventilated with openable sash windows on opposite sides of the building. The floor area is approximately 37.6\,m$^2$ and the floor-to-ceiling height is 2.7\,m; \citet{Song18} provide full details of the monitored space and the monitoring equipment used but it should be noted the monitored office is not of a modern design and is not well-sealed no{r} well-insulated. For monitored data it is worth considering how to appropriately select the ambient \co concentration, $C_0$, since atmospheric levels do vary slightly and \co sensors can exhibit a base-line drift over time. For all our analysis based on monitored \co data we decided to allow the ambient \co concentration to vary taking its value each day to be the mean value observed between 05:00 and 06:00.

\subsubsection{The role of opening windows in reducing risk} \label{sec:resWind}

\begin{figure}
             \begin{center}
             \includegraphics[width=\textwidth]{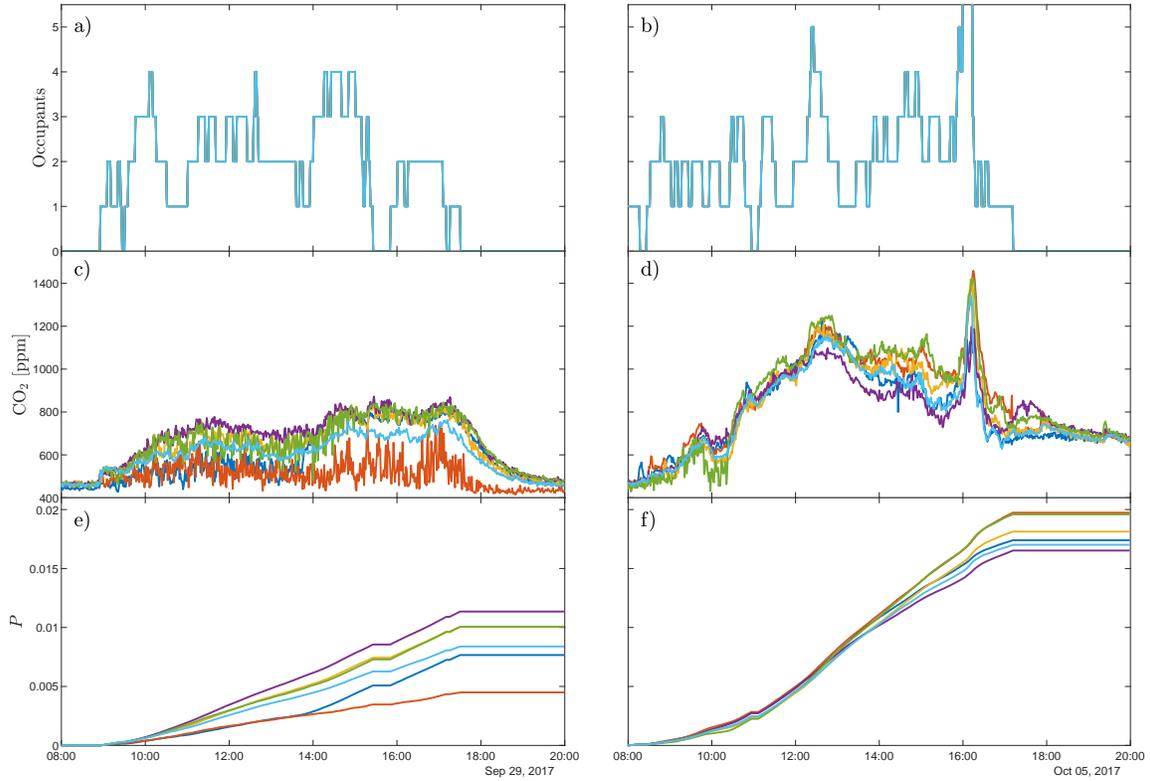}
             \end{center}
             \caption{The intra-day variation in occupancy (upper panes, a) and b)), monitored \co (middle panes, c) and d)) and the corresponding risk of airborne spread of COVID-19 (lower panes, e) and f)) during 29\upth Sep 2017 (left-hand panes, a), c) and e)) and 5\upth Oct 2017 (the right-hand panes, b), d) and f)). Data are plotted from six \co monitors placed at various locations and heights (between 73\,cm and 242\,cm from the floor). On the 29\upth Sep (left-hand panes) windows on opposite sides of the room were opened (creating an opened area of around 0.24\,m$^2$) from 08:00 until 20:00 whilst on the 5\upth Oct (right-hand panes) the windows remained closed all day.}\label{fig:MAG4}
\end{figure}

Figure \ref{fig:MAG4} a) and b) show the occupancy profiles during two days in 2017. During 29\upth Sep the windows were opened on both sides of the building (providing an opened area of 0.24\,m$^2$) at around 09:00 and remained so until after 20:00; whilst on 5\upth Oct the windows remained closed all day and we note that the spike in \co at around 16:15 on this day corresponds to a brief visit during which 22 people were in the office. The monitored \co profiles (figure \ref{fig:MAG4} c) and d)) were obtained at six locations of differing height{s} (between 73\,cm and 242\,cm from the floor) and position{s} within the office. It is most striking that the \co levels are markedly higher on 5\upth Oct when the windows remained closed. Crucially, these elevated \co levels translate into increased risk of airborne infection for the occupants -- in this case the risk of infection being approximately doubled on the day when the windows remained shut. In addition, at times (e.g. between about {14:00} and {17:00} on 25\upth Sep) there is a marked variation in measured \co levels dependent on location. It follows that this variation in \co is reflected in the infection risk levels which indicates that {the location} within the office at which one was breathing affected the risk of infection by around 20\% on the day the windows were closed and a much more substantial variation on the day the windows were open.

\begin{figure}
             \begin{center}
             \includegraphics[width=\textwidth]{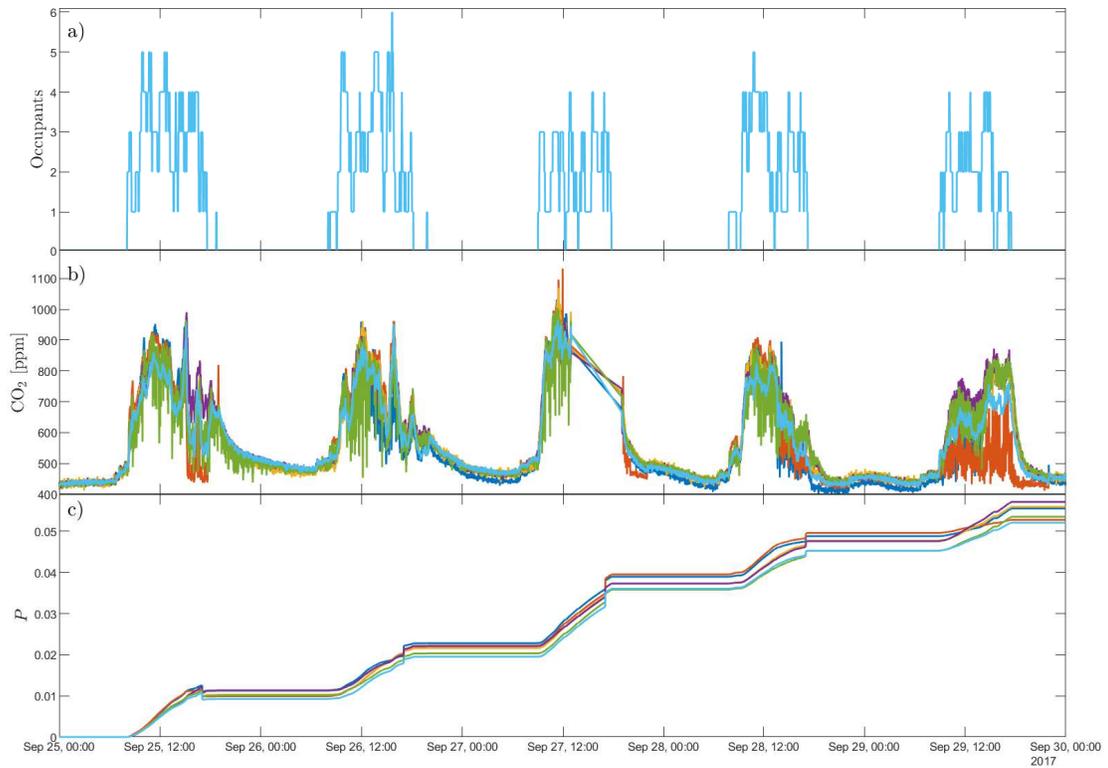}
             \end{center}
             \caption{The variation in a) occupancy, b) monitored \co\!, and c) the corresponding risk of airborne spread of COVID-19, over a period of pre/asymptomatic occupancy of the monitored office. Data are plotted from six \co monitors placed at various locations and heights (between 73\,cm and 242\,cm from the floor).}\label{fig:MAG5days}
\end{figure}

Figure \ref{fig:MAG5days} a) shows the occupancy data for the monitored office over a five day period in September 2017. During this five day period the office windows were open for some significant portion of each day. The accompanying monitored \co data is shown in figure \ref{fig:MAG5days} b) and it should be noted that data are missing between 13:30 and 19:00 on 27\upth September. The risk of airborne infection for COVID-19 is shown in {the} lower-pane with the risk rising gradually over the period of pre/asymptomatic infectivity reaching an absolute risk $0.059 \leq P_A \leq 0.064$ depending on where within the office the occupant would have been located. \edit{We note that two of the sensors were positioned very close to the windows. However, perhaps surprisingly, for the five day period the \co concentrations measured in these positions were not significantly below that measured elsewhere within the office and the airborne infection risk estimated from the sensors near the windows was not systematically lower than those positioned in the centre of the room. One of the sensor placed close to the window, the data marked in orange in figures \ref{fig:MAG4}c), \ref{fig:MAG4}e), \ref{fig:MAG5days}b), and \ref{fig:MAG5days}c) does show levels which are significantly below the other sensors during 29\upth Sep --- however, as figure \ref{fig:MAG5days}c) shows over the five day period (25\upth\!--29\upth Sep) the impact is not drastic with the risk inferred from all sensors lying within $\pm5\%$ of the mean.} The expected number of secondary infections for the monitored office over this period is $0.3 \leq S_I \leq 0.32$ -- reassuringly below unity and indicating that this naturally ventilated office was likely to have been receiving somewhere between 10\,l/s/p and 20\,l/s/p (see table \ref{tab:R}). From examination of the monitored office during periods when the windows were closed the expected number of secondary infections might approximately double to $S_I \approx 0.6$. We note however, that \edit{these risks might be} considerably higher for more modern well-sealed offices.

\edit{
\section{Retrospective modelling using COVID-19 as an example} \label{sec:resRet}

\begin{figure}
             \begin{center}
             \includegraphics[width=0.95\textwidth]{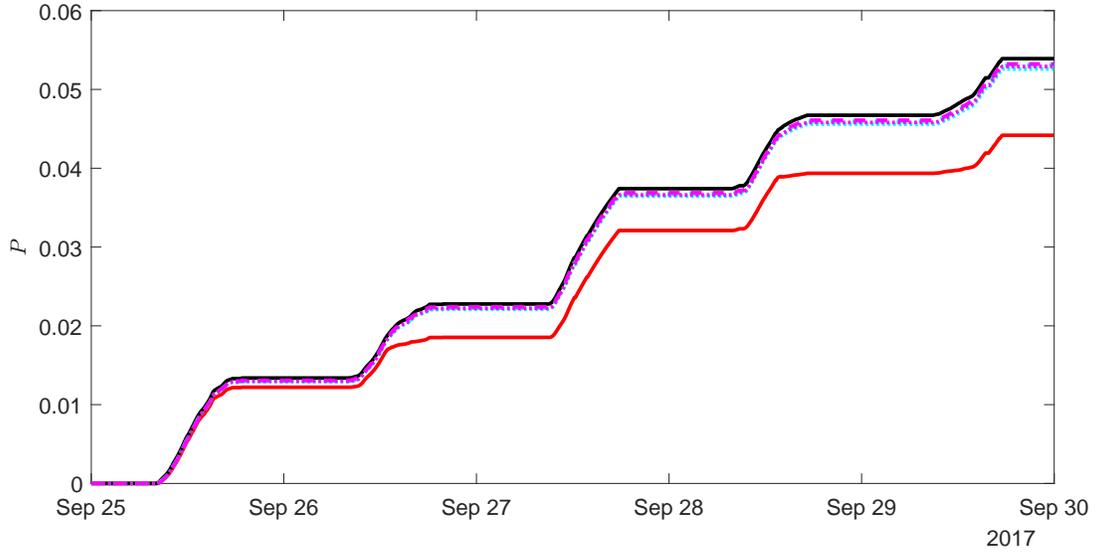}
             \end{center}
             \caption{The variation in the likelihood of airborne infection risk, $P$, over a five day period based on \co data from one sensor within the monitored office. The black curve shows the likelihood based on the predictive methodology (\ref{eq:Pus}), i.e. the same data from this sensor shown in figure \ref{fig:MAG5days}. Estimates of the airborne infection risk, taking the same disease parameterisation and occupancy, but reconstructing the risk using the full equations using the raw \co data are marked with by the red curve. Estimates of the airborne infection risk reconstructed using the full equations and \co data subjected to a low-pass filter are marked with by cyan curves for infinite-duration impulse response filter and magenta curves for the Savitzky-Golay filter. Data are shown for passband frequencies, and filtering windows, based on $T_f = 20$\,min (dotted curves) and $T_f = 240$\,min (dashed curves).}\label{fig:RetFilt}
\end{figure}

In the following section we chose to examine a full pre/asymptomatic period within an open-plan office. However, the methodology we present is applicable to any shared indoor space for durations over which the population of occupants remains the same/similar.  

Where retrospective modelling \resp{of an outbreak} is desired it is natural to assume that the infector's occupancy profile may be known (or at least estimated) and so \resp{invoking} the simplifying assumptions that lead to our predictive model, (\ref{eq:Pus}), \resp{is} likely to be inappropriate. In order to account for bespoke infector occupancy profiles one must return to the general equation for the likelihood of airborne infection, i.e. (\ref{eq:Pgen}), which requires solution of both (\ref{eq:Q}), to determine the ventilating flow though the space, and (\ref{eq:figen}), to determine the amount of infectious airborne material present. This retrospective modelling requires that the ventilating flows are inferred from the monitored \co and that these flows are then utilised to determine the dilution of the airborne infectious material being emitted by any infectors as and when they are present. As such, for this retrospective modelling it is necessary to assume that all of the air within the space is well-mixed. Where indoor concentrations are expected to vary spatially within a single space then multiple \co monitors can be deployed to establish estimates of the errors introduced by the necessity to assume well-mixed air. The \co data presented herein is one such case, a naturally ventilated office with no mechanical means to generate a well-mixed environment. However, despite having sensors positioned both in the centre of the room and near the windows, over the full five day period (figure \ref{fig:MAG5days}) the spatial variation in \co concentration \pfl{alone only changes} the risk inferred from any sensor \pfl{by at most} $\pm5\%$ of the mean; this despite the fact that for certain periods \co concentrations varied considerably between the sensors.

To test the methodology of retrospectively inferring the risk from \co measurements we select the data from one sensor (plotted in green in figures \ref{fig:MAG5days}), which is relatively typical within the set recorded. We note that the results reported hereinafter have been tested for each of the six sensors with no notable differences arising. In order to establish the validity to enable comparison of the results from the retrospective analysis to be compared to the `true' predicted previously. Within figure \ref{fig:RetFilt} the results of the risk from the predictive method are plotted as the black curve resulting in $P_A = 0.054$. However, when the same \co data are used to solve (\ref{eq:Q}) and (\ref{eq:figen}), which are then substituted into (\ref{eq:Pgen}) and integrated, i.e. the retrospective method \resp{(with the proportion of infectors, $I/n$, taken to be constant solely to enable direct comparison to the predictive method), the airborne infection risk (marked by the red curve) is significantly lower with $P_A = 0.044$. This difference arises because the retrospective methods requires discrete observational data to be differentiated. Passing these data, which were recorded every minute, through a low-pass filter reduces much of this difference. For all our data, the airborne infection risk implied by the retrospective method was relatively insensitive to the choice of low-pass filter (we tested the finite-duration impulse response, infinite-duration impulse response (IIR), and a third-order Savitzky-Golay filter in Matlab) and also to the passband frequencies $T_f^{-1}$ with $20$\,min$\leq T_f \leq 240$\,min (or filtering windows of $T_f$ in the case of the Savitzky-Golay filter). 

Figure \ref{fig:RetFilt} also shows the retrospective airborne infection risk inferred from the \co data, with a passband frequency and filtering window set by $T_f = 20$\,min, for an IIR (dotted cyan line) and a Savitzky-Golay filter (dotted magenta line), and with $T_f = 240$\,min for an IIR (dashed cyan line) and a Savitzky-Golay filter (dashed magenta line) --- all of which result in $P_A = 0.053$ (when rounded to two significant figures). As such, we can report that with appropriate filtering, airborne infection risk can be reconstructed retrospectively based on monitored \co data. Based on all available data, the method appears to provide good accuracy when the \co data is subjected to a low-pass filter, irrespective of the choice of filter, and exhibits an adequately low sensitivity to the choice of passband frequency, or filtering window. 

For completeness, each night the \co concentration within the office notionally reached ambient levels and these levels varied slightly from day-to-day, and we observed singularities in the implied flow rate when the denominator of the right hand term of (\ref{eq:Q}), namely $(C-C_0)$, became zero. These could have either been avoided by setting the ambient \co concentration artificially low but that would result in artificially high risk being reported. Instead, we replaced the denominator with $\textrm{max}(C-C_0,10^{-6})$ which both avoided the singularities and did not overstate the risk --- this insertion can be justified upon realising that the tolerance of \co monitors typically far exceeds $10^{-6}$ or 1\,ppm. It is interesting to note that irrespective of the filtering we were never able to obtain estimates of the instantaneous ventilation flow rates, these typically fluctuated unreasonably in the range $\pm 100\,$ACH (air changes per hour). However, since these instantaneous ventilation flow rates only affect the likelihood via (\ref{eq:figen}) which is then integrated twice, with respect to time, such unreasonable fluctuations do not, in practice, significantly affect estimates of airborne infection risk.

\begin{figure}
             \begin{center}
             \includegraphics[width=0.95\textwidth]{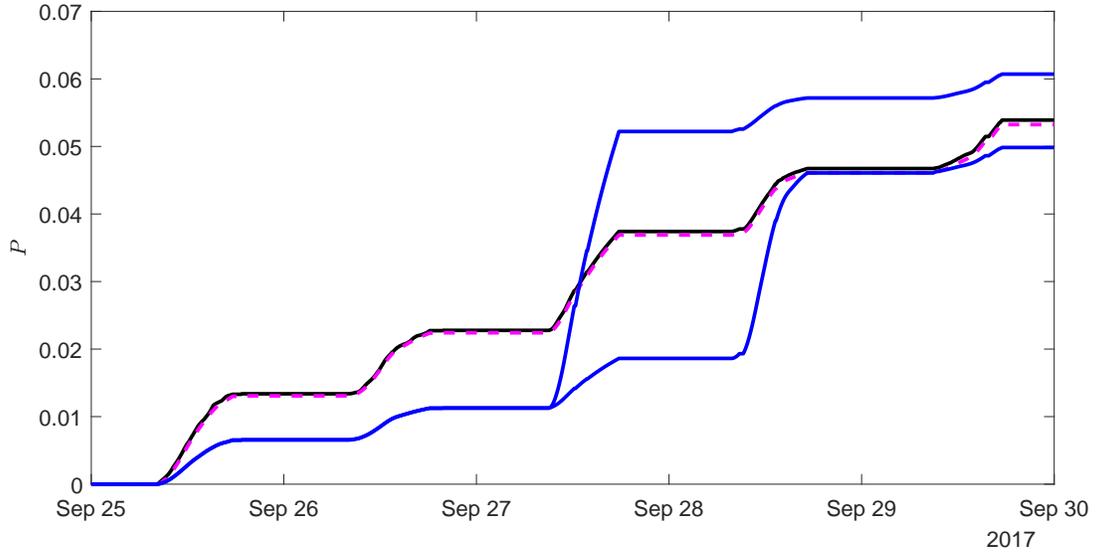}
             \end{center}
             \caption{The variation in the likelihood of airborne infection risk, $P$, over a five day period based on \co data from one sensor within the monitored office. For constant quanta generation rate of $q=1$\,quanta/hr: the black curve shows the likelihood based on the predictive methodology (\ref{eq:Pus}) and the dashed magenta curve the likelihood reconstructed from the low-pass filtered \co data. The blue curves show the results assuming the disease is such that the infectivity on day 3 is greater than on the other four days but for fair comparison we parameterise the disease such that integral quanta over the five day period remains unchanged. In this case we chose values of $q=0.5$\,quanta/hr for the fours days, rising to $q=3$\,quanta/hr on the more infectious day. The plot shows the final risk can be either above (when $q=3$\,quanta/hr on the second day, i.e. 27\upth Sep) or below (when $q=3$\,quanta/hr on the third day, i.e. 28\upth Sep) the results for the constant quanta case; with this being simply determined by whether rebreathed air was more prevalent (than the average over the full duration, here five days) during the more infectious period, or not.}\label{fig:RetVarq}
\end{figure}

In order to demonstrate the flexibility of this method, we could examine the effects on the likelihood of airborne infection of a particular arrival and departure schedule of an infector during the five day period. However, what is perhaps more interesting is to consider a case in which the infectivity of the disease varies during the pre/asympomatic period, as is thought to be the case for COVID-19. We parameterise this variation in infectivity  via the quanta generation rate $q(t)$, and for reasonable comparison we ensure that the intergral of $q(t)$ over the five days remains unchanged. Consider the case that on day 3 the disease was particularly infectious with $q(t)=3$\,quanta/hr, while selecting $q(t)=0.5$\,quanta/hr for the other fours days, thus ensuring that an average of $q=1$\,quanta/hr. In order to properly account for varied quanta generation rates the subject of the integral in (\ref{eq:Pgen}) needs to be evaluated via
\begin{equation}
    \frac{\ud f_i  \, q}{\ud t} = \frac{I \, p \, q}{V} - \frac{f_i \, q \; Q}{V} \, . \label{eq:fiqgen}
\end{equation}
The results are shown in figure \ref{fig:RetVarq}. For reference the likelihood of airborne infection with constant $q=1$\,quanta/hr are shown for the predictive method, by the black curve, and the retrospective method (based on low-pass filtered \co data), by the dashed magenta curve \resp{--- both assuming $I/n$ is constant, purely to enable comparison}. Two blue curves mark the results for the case of variable infectivity being considered. For one, the more infectious day was taken to be the third day (27\upth Sep), and for the other in was taken to be the forth day (28\upth Sep). Both blue curves initially indicate lower risk, for the days on which $q(t)=0.5$\,quanta/hr, and then show marked increases in the risk on the more infectious day before continuing to increase (at a rate less than that of the constant quanta generation rate). The risk for the case where the disease was more infectious on the 27\upth Sep finishes significantly above the risk of the constant quanta generation rate case; the opposite is true for the case where disease was assumed more infectious on the 28\upth Sep. This is simply because the integral \co levels during 27\upth Sep were above the average for the five days and those on the 28\upth Sep were below the average. This is in line with expectations based on inspection of the system of equations since the risk is determined by the integral of $f_i \, q$ over time. Crucially, these results provide successful demonstration of the principles of retrospectively evaluating the airborne infection risk for a general case based on actual \co measurements.
}}
\section{Conclusion} \label{sec:conc}

Taking COVID-19 as an example \edit{relatively simple models} have been derived to estimate the likelihood of airborne infection within indoor spaces which \edit{can account for variable occupancy levels, bespoke infector behaviours, and diseases for which the infectivity varies in time during the infectious period}. Our model\edit{s} require only monitored or modelled data for \co\!, occupancy \& infector levels, and estimates of appropriate quanta generation rates \edit{(which in the most general case can vary in time)}. A modelled office and, separately, monitored data were used to demonstrate results. 

We conclude that for open-plan offices, regularly attended by the same/simliar people, which have ventilation provision in-line with UK guidance \citep[e.g.][]{CIBSEA} \edit{then attendance of quiet desk-based} work is unlikely to significantly contribute to the spread of COVID-19 \edit{via the airborne route}. However, \resp{this changes} should these spaces be poorly ventilated (e.g. 4\,l/p/s) \edit{since then, through a single infected person attending the office, the expected number of people becoming infected via the airborne route is close to unity\resp{, and estimates rise well above for the SARS-CoV-2 variant B1.1.7.}}. Even for adequately ventilated spaces if the occupants are very vocal (e.g. a call-centre) then, in the presence of a single infector, one could expect attendance of the office to give rise to \resp{between} two \resp{and four} new COVID-19 infections\edit{, and so for these spaces we conclude that attendance could significantly contribute to the spread of COVID-19 via the airborne route}. \edit{Due to their ill-defined nature, we choose not to focus on superspreader scenarios but instead we primarily considered conditions more in-line with the median of the population.} Should we have chosen to examine superspreaders then our results would have been more alarming. 

\resp{Crucially, the above conclusions highlight that just by making readily achievable changes to the indoor environment can alter whether, or not, the attendance of work within an open-plan office might significantly contribute to the spread of COVID via the airborne route. These changes include, altering ventilation between rates that might actually be supplied to working offices, and changes in behaviours that can be expected to occur with different office usage. More successful variants of the virus, e.g. B1.1.7, may only result in a greater number of the environmental scenarios giving rise to significant numbers of secondary infections. This highlights that as, and when, communal workplace practices are re-established there is a need to better mitigate against the airborne spread of COVID-19, while maintaining every effort to reduce the spread by other transmission routes.}

\edit{Assessing and maintaining existing ventilation provision is the primary step in understanding the mitigation needs within an indoor space against for the airborne spread of COVID-19. To that end, we} recommend that more widespread monitoring of \co is carried out within occupied spaces. \edit{Doing so will provide a step towards practically assessing the actual ventilation provision being supplied to these spaces.} Where these spaces can be considered to broadly conform to our definition of a regularly attended space then we further recommend that occupancy profiles are recorded. In so doing, we provide a simple methodology \edit{with which to calculate the expected number of secondary COVID-19 infections arising, via the airborne route, within the monitored space}. Irrespective of this, we believe that an indication of the rate of increase in infection risk (see (\ref{eq:Pinc})) should be consider{ed} for all indoor spaces; \edit{this can expressed for the general case as $ \lambda (1-P) = f_{i} \, q (1-P)$. In cases where estimates of the likelihood $P$ area being recorded then the term, $(1-P)$, can be included. Otherwise, a `worst case' estimate can be easily obtained by taking the rate of increase in infection risk to be $f_{i} \, q$. However, there is no easily measurable proxy for $f_{i} \, q$ but where some predictive measure is required then an estimate can be obtained from consideration of our model, (\ref{eq:Pus}), giving $\lambda = (C - C_0)\, \alpha \, q / C_a $. For any given disease and chosen indoor space this shows that excess \co determines the rate at which airborne infection risk is increasing, i.e. $\lambda \propto (C - C_0) $ with all other variables being independent of environmental conditions. Hence monitoring the excess \co within spaces, for which occupants are enabled to make appropriate change, may be of significant benefit in mitigating airborne infection risk}. 

\edit{From a practical perspective it may be challenging to increase ventilation provision without significant time and investment, or without compromising occupants thermal comfort (which risks causing unwanted interventions). However, the formulation of our model makes it simple to demonstrate that reducing occupancy by a factor $r$ and keeping the ventilation provision unchanged reduces the expected number of secondary infections by a factor $r^2$ for all of the scenarios considered. This observation may aid the safer re-establishment of open-plan offices where partial occupancy is of benefit and their environmental design is appropriate then introducing week-in week-out working may result in tolerable airborne infection risks for COVID-19 while offering benefits to both employees and employers.}

Finally, we conclude that our model indicates that the risk of COVID-19 being spread by the airborne route is not insignificant and varies widely with activity level and environmental conditions which are predominantly determined by the bulk supply of outdoor air. 


\bibliographystyle{jfmF} 
\bibliography{bibl_USE.bib}

\appendix
\section{Modelling from first principles} \label{app:first}

In order to demonstrate the underlying assumptions and highlight the limits of applicability we revisit the formulation of the Wells-Riley equation \citep{Wells55,Riley78}. We wish to determine the likelihood, $P$, that infection spreads within a given indoor space during a time interval $T$. Denoting the probability that no one becomes infected during this time $P(0,T)$ gives that
\begin{equation}
    P = 1 - P(0,T) \, . \label{eq:P}
\end{equation}
The number of infected people, $I$, is discrete (i.e. an integer) and since time is continuous we can consider a small period of time, $\delta t$, during which either no one becomes infected or one person becomes infected. Defining the infectivity rate $\lambda$ (see below for a detailed discussion) gives the likelihood that one person becomes infected during this small time period as
\begin{equation}
    P(1,\delta t) = \lambda \, \delta t \, .
\end{equation}
Assuming that each infection occurs independently of the last
\begin{equation}
    P(0,t + \delta t) = P(0,t)[1 - P(1,\delta t)] = P(0,t)[1 - \lambda \, \delta t] \, .
\end{equation}
Rearranging and taking the limit $\delta t \to 0$ gives
\begin{equation}
    \frac{P(0,t + \delta t) - P(0,t)}{\delta t} \equiv \frac{\ud P(0,t)}{\ud t} = -\lambda P(0,t) \, , \label{eq:Pinc}
\end{equation}
integrating and substituting into (\ref{eq:P}) gives
\begin{equation}
    P = 1 - \exp{\left( - \int_0^T \lambda \, \ud t \right) } \, . 
\end{equation}
Hence the likelihood $P$ can be evaluated for any known functional form of $\lambda$, crucially, as we go on to demonstrate, this can inferred from data measured within indoor spaces which records occupancy level profiles and \co concentrations. 

\section{The challenges of measuring ventilation rates or inferring ventilation rates from monitored \co} \label{app:Q}

The original formulation of the Wells-Riley equation requires parameterisation of not only the quanta generation rate, $q$, but also evaluation of the infectivity rate which in the general case must be evaluated via the integral
\begin{equation}
    \int_{0}^{T} \frac{I\, p \, q}{Q} \; \ud t \, . \label{eq:int_WR}
\end{equation}
Assuming broadly constant activity levels within a space the breathing rate, $p$, and (for a given disease) the quanta generation rate can be regarded as time independent. Furthermore it may be reasonable to assume the number of infected people within space remains unchanged if one is examining the likelihood of spread. Moreover, if the ventilation rate $Q$ can be assumed constant and if the space is in steady-state then the probability of infection occurring during the period $T$ can be simply expressed as
\begin{equation}
    P = 1 - \ex \left( -\frac{I\,p\,q}{Q}T \right) \,. \label{eq:WR}
\end{equation}
However, it is this last assumption that is most troubling since it is only reasonable if for a time exceeding the transient ventilation effects (which typically remain significant for multiple hours): the opened area of all connections to the space (windows, doors, vents, etc...) remain unchanged, infiltration rates remain constant (or negligible), and the outdoor air supply rates are insensitive to changes that arise in the pressure differences between indoors and outdoors due to changes in temperature and wind --- this makes application of (\ref{eq:WR}) difficult. As we will discuss, measuring or inferring the ventilation rate within an operational occupied space is non trivial and hence evaluating (\ref{eq:int_WR}) is impractical. The insightful work of \citet{Rudnick03} solved many of these challenges in assessing airborne infection risk and forms the basis for the modelling described in \S \ref{sec:model}.

The magnitude of infection risk changes with the seasons for numerous viral infections (including influenza) and these may  arise for a variety of factors. These might include: changes in the viability of the virus if typical temperatures and/or humidity of indoor environments vary with the season, or if changing levels of natural UV light are significant and effect viability; changes in behaviour, for example, staying indoors more during colder seasons; changes due to the seasonal variations that occur in immunity \citep{Dopico15}; and, crucially for the airborne transmission route changes in ventilation (i.e. outdoor air supply) rates that occur as moderated indoor temperatures are demanded once outdoor temperatures vary with the season. We assert that this last factor, changing outdoor air supply rates, is highly significant yet it is poorly evidenced.

Outdoor air may enter an indoor space via a ventilation system, windows, doors, vents, cracks in the building fabric or, indeed, though the very fabric itself (i.e. many building materials, e.g. bricks, are porous). As such, there is significant scope for both intentional and unintended supply of outdoor air. Directly measuring the air flow through all of the potential pathways for any given indoor space in impractical. Pressure testing can be used to measure infiltration rates but cannot assess the ventilation rates in operational settings. 

Indoors, human activity is typically the major source of \co while outdoor \co levels remain broadly constant. Therefore, if the rate of \co production from human activity within a space can be estimated, \co provides a suitable proxy from which to attempt inference of the ventilation/outdoor air supply rate within the space. Consider an indoor space in which both occupancy levels and \co are monitored. If the activity levels of individuals remains broadly similar and the \co monitored, $C$, can be regarded as indicative of the \co levels throughout the space, i.e. the \co is relatively well-mixed within the indoor air (note that only point measurements of \co are practically possible), then since \co is inert its conservation requires that 
\begin{equation}
    V \frac{\ud C}{\ud t} = n \, p \, C_{a} - Q \, (C - C_0) \, , \label{eq:cons}
\end{equation}
where $n$ is the number of people in the space and $C_a$ is the volume fraction of \co added to exhaled breath during breathing. As discussed, it is unwise to regard the ventilation rate as constant for most indoor spaces, i.e. $Q=Q(t)$, and this renders (\ref{eq:cons}) non trivial to integrate analytically. Thus if one wishes to examine the ventilation rate
\begin{equation}
    Q  = n \, p \, \frac{C_{a}}{(C - C_0)} - \frac{V}{(C - C_0)} \frac{\ud C}{\ud t}  \, , \label{eq:Q} \, 
\end{equation}
can be evaluated with monitored occupancy and \co data\resp{, assuming a reasonable estimate of individuals \co flux, $p \, C_{a}$, can be made (which in practice varies, in particular, with age, gender, and activity levels)}. However, as with all real-world data, the \co signal is likely to contain some non-negligible level of noise and the dependence on $\ud C/\ud t$ will, in most cases, render evaluation of ventilation/outdoor air supply rate via (\ref{eq:Q}) unsuitable.

An alternate approach is to examine the monitored occupancy data to determine the time at which the room becomes unoccupied. Assuming that the ventilation rate remains unchanged thereafter (which will only be the case if ventilation systems are left operational, and any changes in the ventilation/outdoor air supply rate due to the effects of wind and temperature variations are negligible) the \co concentration within the space, assuming the air within remains relatively well-mixed, will decay exponentially. By exponential fitting to the monitored data during this period a ventilation/outdoor air supply rate can be inferred. However, curve fitting to real-world data is prone to variability due to choices of the input parameters (e.g. the period over which exponential decay to determine to be observed) and subject to influence from noise within the data, thereby rendering the results unreliable. Moreover, this process is hard to automate and so typically requires significant manual intervention, making it unsuitable for the analysis of large data sets.    

\vspace{\baselineskip}

In summary, direct or inferred measurements of ventilation/outdoor air supply rates are extremely challenging. For this reason, and those described in \S \ref{sec:model}, we council that to assess airborne infection risks no attempts be made to directly assess outdoor air supply/ventilation rates to indoor spaces. Instead, we suggest widespread monitoring of \co within spaces combined with measured/estimated occupancy profiles, which with application of our extensions (\S \ref{sec:model}) to the work of \citet{Rudnick03} can be used to directly assess the airborne infection risk within a given space. Where required, simple modelling can be carried out to inform and assess practical mitigation strategies.
\vspace{\baselineskip}

\section*{Author Contributions}

We all work hard and produced this article together.

\subsection*{Acknowledgements}
{This work was undertaken as a contribution to the Rapid Assistance in Modelling the Pandemic (RAMP) initiative, coordinated by the Royal Society, and  was supported by the UK Engineering and Physical Sciences Research Council (EPSRC) Grand Challenge grant ‘Managing Air for Green Inner Cities' (MAGIC) grant number EP/N010221/1.} HCB acknowledges insightful conversations with Dr Marco-Felipe King, and the debugging support of Carolanne Vouriot. 


\subsection*{Competing interests}

The authors declare no competing interests.

\end{document}